# The fast and the slow axonal transport: a unified approach based on cargo and molecular motors coupled dynamics.

Alexandre Y. C. Cho, Victor R. C. M. Roque, Carla Goldman

Universidade de Sao Paulo, Instituto de Fisica

Rua do Matao, 1371

05508-090 Sao Paulo, SP, Brazil

May 2020

## Abstract

The origins of the large differences observed to the rates with which the diverse particles are conveyed along axonal microtubules are still a matter of debate in the literature. There is evidence that certain neurodegenerative diseases may be triggered by disturbances to the related transport processes. Motivated by this, we employ a model to investigate the mobility properties of certain cargoes which dynamics are coupled with that of molecular motors on crowded microtubules. For certain initial and boundary conditions, we use the method of characteristics to resolve perturbatively the pair of equations of the Burgers type resulting from a mean-field approach to the original microscopic stochastic model. Extensions to the non-perturbative limits are explored numerically. In this context, we were able to figure out conditions under which cargos average velocities may differ up to orders of magnitude just by changing the number of motors on the considered track. We then discuss possibilities to connect these theoretical predictions with available experimental data about axon transport.

# 1  Introduction

The diverse types of particles that are usually transported along axons can be grouped into two major categories characterized by the average speed $v$ at which this transport is put into effect. The group of fast-moving particles with $v \sim 0.5-5 \ \mu m/s$ comprises membranous organelles as the Golgy-derived vesicles, mitochondria, endosomes and lysosomes, among others. The groups of slow-moving particles with $v \sim 0.003-0.03 \ \mu m/s$ for slow component-**a** (Sc**a**) and $v \sim 0.02-0.09 \ \mu m/s$ for slow component-**b** (Sc**b**) comprise nonmembranous neurofilaments, cytoskeletal and cytosolic proteins, among others.

The general biological interest of related studies relies on the fact that certain neurological diseases are believed to be directly associated with failures of the system to keep the transport at the right rates leading, eventually, to local particle accumulation [1]. The review of S. Roy [2] offers an historical account of experimental achievements in this field and of the major hypotheses in the literature to explain the huge differences in the rates of particle transport along axons. One of the models used in these studies, the *Dynamic Recruitment model* proposed to explain the slow movement, requires the presence of a common *carrier structure* [3],[4] or even fast moving cargo vesicles [5] onto which cytosolic proteins would transiently get attached to. The slow rate with which these proteins are conveyed along the axon would then be attributed to the relative large intervals of time along which they stay detached from the carrier [6]. A stochastic model accounting for these ideas is considered in [7]. The other major hypothesis, the *Stop and Go model*, attributes the causes of slow movement to the ability of neurofilaments, for example, to pause during transit. The duration of such pauses would be determinant to the resulting slow rates, although their origins and eventual mechanisms of control are not clarified yet.

Evidences indicating that individual Cytoskeletal polymers conveyed by slow transport in axons can also move as fast as membranous organelles suggest that both fast and slow transport in axons may be due to a single mechanism. Actually, a unified view of axonal transport had already been proposed by Ochs [8] according to which the only actual directed movement would be that of fast cargoes, the detected slow movement would be due to local and casual rearrangements of particles, not resulting in long-range transport. In line with this it is argued these days that such unified mechanism is supported by the dynamics of molecular motors and on their ability to transport a variety of particles, referred generically as cargoes, as they move on structured "tracks", i.e. along axonal microtubules [9]. Yet, one still may question about the specific motors that would drive cargoes at such slow rates as shown by Sc**a** and Sc**b** [1]. In view of this, we believe that although the description based on molecular motors seems promising, it requires to be complemented by mechanisms that regulate cargo-motor interactions, able to explain such huge differences between the rates of slow and fast transport in the absence of a putative carrier. Our intention here is to contribute in this regard.

To that end, we investigate specific properties of a stochastic model proposed elsewhere [10], [11], [12]. It is conceived on the idea that the transport of cargoes by molecular motors is based on a mechanism of *cargo hopping among motors* which, in turn, depends crucially on the quantity of motor available on the track. Here, we seek for quantitative predictions for the cargo average velocities under these conditions. We explore situations under which motor jamming on the track may promote cargo transport.

We should remark that there are always two main aspects concerning the dependence of cargo transport on molecular motors in such systems. One of these is related to the microscopic mechanisms of motor-cargo binding affinity. The other concerns the dependence of the transport on the number of available motors on track, not only on the type of motor directly bound to each cargo [13], [14], [15]. In order to investigate the dependence on motor occupation, we base our quantitative analysis on a model that describes the microscopic dynamics of motors and cargoes coupled through stochastic processes of ASEP type (Asymmetric Simple Exclusion Process). The model allows for two microscopic mechanisms for cargo movement. One of these describes the movement of cargo while bound to a motor; the other describes cargo hopping between pairs of neighbor motors on a track. There is only one type of motor with bias in a definite direction; without prejudice for the analysis, this is chosen anterograde. We have already studied possibilities to observe bidirectional movement in similar model systems [10],[11],[12]. The main concern of the present study is related to quantitative estimations for the average velocities developed by cargoes subject to these two mechanisms. In this context, we show the conditions under which the average cargo speed may vary up to orders of magnitude simply adjusting the amount of motor available on axons. This opens a new possibility to understand the observed differences in the rates of transport depicted by real axons in quantitative terms and under a unified perspective.

In Section 2 we derive the mean-field equations for the model. The resulting non-linear coupled equations of Burgers type is studied analytically in Section 3 for certain initial and boundary conditions using the Method of Characteristics. To our knowledge, such solutions to this particular system of equations represent an original contribution. As we shall show, these are achieved under a perturbative approximation. The analytical expressions obtained in this way pointed out to the relevant regions of parameters, and particular initial and boundary values illustrating the role of each mechanism in producing such differences to the average cargo velocities, as observed. For more general conditions, away from the perturbative limits, we base our analysis on a numerical study to resolve the particular system of non-linear equations of interest [16]. This is explained in Section 4. We discuss our results in Section 5. Some conclusions and additional remarks are in Section 6.

## 2 The cargo hopping model

The kind of ASEP model we shall consider to describe the stochastic dynamics of molecular motors and cargoes has already been proposed elsewhere to study the so called bidirectional movement [10],[11],[12]. The model version analyzed here embraces both mechanisms of cargo transport at the microscopic level namely, the isolated motion of a cargo attached to a motor and also the mechanism of cargo hopping from motor to motor. The idea is to circumvent difficulties for long-range cargo transport due to motor jamming (congestion) on axons [17]. Accordingly, an axonal track is represented by a one-dimensional lattice which sites may be occupied at each time either by an unloaded motor or by a loaded motor or else it can be empty. The model is defined by the dynamics of occupation, as follows:

$$
\begin{array}{rlll}
\textbf{(a)} & 10 & \to & 01 \quad \text{with probability } p \\
\textbf{(b)} & 20 & \to & 02 \quad \text{with probability } q \\
\textbf{(c)} & 12 & \to & 21 \quad \text{with probability } f \\
\textbf{(d)} & 21 & \to & 12 \quad \text{with probability } g
\end{array}
\qquad (1)
$$

Label 0 is assigned to an empty site. Label 1 is assigned to a site that is occupied by a motor carrying no cargo - an unloaded motor. Label 2 is assigned to a site occupied by a motor attached to a cargo - a loaded motor. The pairs shown in (1) indicate that each of the possible processes is represented by the occupation of two neighboring sites. Parameters $p, q, f, g$ are the probabilities for the occurrence of each process per time interval $(\Delta t)$. Process (a) represents an elementary movement of a biased motor from cell body towards axon´s terminal (anterograde direction). Process (b) represents an elementary process of the kind considered in (a) of a motor when carrying a cargo particle. The remaining processes describe the exchanging of cargoes between neighboring motors, either to the left (c) or to the right (d). A condition for processes (a) and (b) to take place is that the site at the right of the motor remains empty during the time interval $\Delta t$. The two processes of exchanging depend on the presence of a free motor at the right (c) or at the left (d) of the one attached to the cargo.

## 2.1 The continuous limit and mean field approximation.

In order to obtain the mean-field equations corresponding to processes (1 a-d) for continuous-time and space intervals, we follow the procedure described in [18] that nevertheless accounts only for one type of particle. In order to extend it to take into account the two types of particles present in our model, we define the vector:

$$P^t[n_1, n_2, n_3, ...n_i, n_{i+1}, ..., n_{i+k}, ..., n_N]. \tag{2}$$

For any $j = 1,..N$, $n_j = \{0, 1, 2\}$, it represents the probability of any configuration of particles and holes occupying the sites of the one-dimensional lattice at time $t$. The marginal probabilities are defined accordingly:

$$P_i^t[n_i, n_{i+1}, ..., n_{i+k}] = \sum_{\substack{\{n_j\} \\ (1 \leq j < i;\ i+k < j \leq N)}} P^t[n_1, n_2, n_3, ...n_i, n_{i+1}, ..., n_{i+k}, ..., n_N] \tag{3}$$

where $N$ is the total number of lattice sites. With this, one represents the average densities of particles 1 and 2 at each site $i$ by:

$$\rho_i^t[1] = P_i^t[1]/N$$

$$\rho_i^t[2] = P_i^t[2]/N \tag{4}$$

The temporal behavior of the density at any site $i$ can now be evaluated by relating it to the incoming flux of particles into that site, observing the processes defined through (1), as follows:

$$\frac{1}{\Delta t}(P_i^{t+\Delta t}[1] - P_i^t[1]) = \frac{1}{\Delta t}\{pP_{i-1}^t[10] - pP_i^t[10] + \tag{5}$$

$$+ gP_i^t[21] - gP_{i-1}^t[21] + fP_{i-1}^t[12] - fP_i^t[12]\}$$

and

$$\frac{1}{\Delta t}(P_i^{t+\Delta t}[2] - P_i^t[2]) = \frac{1}{\Delta t}\{qP_{i-1}^t[20] - qP_i^t[20] + \tag{6}$$

$$+ gP_{i-1}^t[21] - gP_i^t[21] + fP_i^t[12] - fP_{i-1}^t[12]\}$$

The mean-field approximation consists in neglecting correlations by taking, for example

$$P_{i-1}^t[20] = \left(P_{i-1}^t[2]P_i^t[0]\right)/N$$

$$P_i^t[21] = \left(P_i^t[2]P_{i+1}^t[1]\right)/N \tag{7}$$

and analogous expressions for the remaining two-point probability vectors. With these, one writes

$$\frac{1}{\Delta t}(P_i^{t+\Delta t}[1] - P_i^t[1]) = \frac{l}{\Delta t}\left[-\frac{p}{Nl}\Delta_i(P_i^t[1]P_{i+1}^t[0]) + \frac{g}{Nl}\Delta_i(P_i^t[2]P_{i+1}^t[1]) - \frac{f}{Nl}\Delta_i(P_i^t[1]P_{i+1}^t[2])\right]$$

$$\frac{1}{\Delta t}(P_i^{t+\Delta t}[2] - P_i^t[2]) = \frac{l}{\Delta t}\left[-\frac{q}{Nl}\Delta_i(P_i^t[2]P_{i+1}^t[0]) + \frac{f}{Nl}\Delta_i(P_i^t[1]P_{i+1}^t[2]) - \frac{g}{Nl}\Delta_i(P_i^t[2]P_{i+1}^t[1])\right] \tag{8}$$

where the notation $\Delta_i$ at the right-hand side of the above equation indicates the difference between the product inside the parenthesis with an analogous obtained by shifting $i \to i-1$. Using that $P_i^t[0] = 1 - P_i^t[1] - P_i^t[2]$ and taking the continuous limit for which both the length $l$ of the lattice sites and $\Delta t \to 0$, the two expressions become,

$$\frac{\partial u}{\partial t} = -\alpha \frac{\partial}{\partial z}[u(1-u) - \theta(uc)]$$

$$\frac{\partial c}{\partial t} = -\beta \frac{\partial}{\partial z}[c(1-c) - \eta(uc)] \tag{9}$$

This system of quasi-linear hyperbolic equations describes the coupled dynamics of both quantities of interest, namely the local density of free motors (i.e., motors that are not attached to cargoes) defined accordingly as $u(z,t) = \lim_{il \to z, l \to 0, N \to \infty} P_i^t[1]/N$, and the local density of motors transiently attached to cargoes $c(z,t) = \lim_{il \to z, l \to 0, N \to \infty} P_i^t[2]/N$, at each time $t$ and position $z$ of the axon. From now on, we shall refer to $c(z,t)$ simply as the cargo density and to $u(z,t)$ as the motor density. The parameters $\alpha, \beta, \theta, \eta$ introduced above are defined in terms of parameters of the original stochastic (microscopic) model as:

$$\alpha = p\xi \qquad \beta = q\xi$$

$$\theta = \frac{p - (f-g)}{p} \qquad \eta = \frac{q + (f-g)}{q} \tag{10}$$

where $\xi = \lim_{\Delta t \to 0; l \to 0} l/\Delta t$ is a scale.

# 3 Analytical solutions: combining a perturbative approach with the method of characteristics

The system (9) can be expressed in a vectorial form,

$$\frac{\partial U(z,t)}{\partial t} + A(U(z,t))\frac{\partial U(z,t)}{\partial z} = 0. \tag{11}$$

where the vector $U$ and the matrix $A$ are defined by:

$$U = \begin{pmatrix} u(z,t) \\ c(z,t) \end{pmatrix} \tag{12}$$

$$A = A(U) = \begin{pmatrix} \alpha(1 - 2u - \theta c) & -\alpha\theta u \\ -\beta\eta c & \beta(1 - 2c - \eta u) \end{pmatrix}. \tag{13}$$

In order to proceed into a quantitative analysis of the equation (11), we examine the case $f = g$, for which $\theta = \eta = 1$. In addition, we parameterize the constants $\alpha$ and $\beta$ related to the hopping probabilities $p$ and $q$, as

$$\alpha = \gamma + \varepsilon \quad \text{and} \quad \beta = \gamma - \varepsilon \tag{14}$$

We may then express the eigenvalues $\lambda_1$ and $\lambda_2$ of $A$ in terms of constants $\gamma$ and $\varepsilon$ which become, for $\varepsilon/\gamma \ll 1$,

$$\lambda_1 \simeq (1 - 2u - 2c)\gamma\left[1 + \frac{\varepsilon}{\gamma}\left(\frac{u-c}{u+c}\right)\right]$$

$$\lambda_2 \simeq (1 - u - c)\gamma\left[1 - \frac{\varepsilon}{\gamma}\left(\frac{u-c}{u+c}\right)\right] \tag{15}$$

Since $\lambda_1$ and $\lambda_2$ are distinct of each other and are both real, the system (9) is strictly hyperbolic. $\lambda_1$ and $\lambda_2$ are related to the velocities of the travelling wave solutions to Eq. (11). The corresponding left eigenvectors $\vec{q}_1$ and $\vec{q}_2$ are calculated perturbatively up to first order in the small parameter $\varepsilon/\gamma$, resulting:

$$\vec{q}_1 \simeq \left(-1, \ \left[-1 - \frac{2\varepsilon}{\gamma}\left(\frac{1-u-c}{u+c}\right)\right]\right)$$

$$\vec{q}_2 \simeq \left(1, \ \left[-\frac{u}{c} + \frac{2\varepsilon}{\gamma}\frac{u}{c}\left(\frac{1-2u-2c}{u+c}\right)\right]\right) \tag{16}$$

The Riemann Invariants $R_1(U)$ and $R_2(U)$ are scalar quantities satisfying $\vec{\nabla}R_i(U) = \vec{q}_i(U)$, $i = \{1, 2\}$ [19], [20]. These quantities are conserved by the dynamics along the set of corresponding characteristic curves $z_1(t)$ and $z_2(t)$ defined by $\partial_t z_1 = \lambda_1$ and $\partial_t z_2 = \lambda_2$. That is $\lambda_1$ and $\lambda_2$ are the instantaneous velocities at each point of the characteristics on the $(z,t)$ plane. In fact,

$$q_i \cdot \left(\frac{\partial U}{\partial t} + \lambda_i \frac{\partial U}{\partial z}\right) = \left(\vec{\nabla} R_i(U)\right) \cdot \left(\frac{\partial U}{\partial t} + \lambda_i \frac{\partial U}{\partial z}\right) = \left(\frac{\partial R_i}{\partial t} + \lambda_i \frac{\partial R_i}{\partial z}\right) \equiv \frac{dR_i}{d\phi_i} = 0 \quad (17)$$

where we have used (11). With regard to expressions (15), one observes that if both $u$ and $c$ are constants along the characteristic curves, then both $\lambda_1$ and $\lambda_2$ are also constants so that the characteristics are straight lines. As indicated in the definition above, the derivative $\frac{d}{d\phi_i}$ is taken along each of the corresponding characteristic curves. Therefore, in order to find $R_i$, $i = \{1,2\}$ the expressions in (16) shall be inserted into the LHS of the first equality in the expression above in order to find a general solution $R_i$ by integrating $\frac{dR_i}{d\phi_i} = 0$.

For $q_1$, it results:

$$-\left(\frac{\partial u}{\partial t} + \lambda_1 \frac{\partial u}{\partial z}\right) - \left(\frac{\partial c}{\partial t} + \lambda_1 \frac{\partial c}{\partial z}\right) - \frac{2\varepsilon}{\gamma} \frac{1-u-c}{u+c}\left(\frac{\partial c}{\partial t} + \lambda_1 \frac{\partial c}{\partial z}\right) = 0 \quad (18)$$

which can be rewritten as

$$\frac{d}{d\phi_1}\left[(1-u-c) - \ln(1-u-c) + \frac{2\varepsilon}{\gamma}c\right] = 0 \quad (19)$$

where the derivative is taken along the characteristics: $\frac{d}{d\phi_1} = \frac{\partial}{\partial t} + \lambda_1 \frac{\partial}{\partial z}$. This means that

$$R_1 = (1-u-c) - \ln(1-u-c) + \frac{2\varepsilon}{\gamma}c \quad (20)$$

is the conserved quantity along the set of characteristic curves associated to $q_1(u,c)$.

For $q_2$, it results:

$$c\left(\frac{\partial (u/c)}{\partial t} - \lambda_2 \frac{\partial (u/c)}{\partial z}\right) + \frac{2\varepsilon}{\gamma}\frac{u}{c}\frac{1-2(u+c)}{u+c}\left(\frac{\partial c}{\partial t} + \lambda_2 \frac{\partial c}{\partial z}\right) = 0. \quad (21)$$

In order to write this in a form analogous to (19) we examine the region of densities for which:

$$u(z,t) = 1/4 + x(z,t)$$

$$c(z,t) = 1/4 + y(z,t) \quad (22)$$

with $|x(z,t)|, |y(z,t)| \ll 1/4$ representing respectively, the excess (or depletion) of motors and cargoes at each point of the axon with respect to a considered density "background" set at $1/4$. With these, Eq. (21) can be approximated by

$$\frac{d}{d\phi_2}\left(\frac{x+1/4}{y+1/4}\right) \simeq 0 \quad (23)$$

where $\dfrac{d}{d\phi_2} \equiv \dfrac{\partial}{\partial t} + \lambda_2 \dfrac{\partial}{\partial z}$. This yields:

$$R_2 \simeq \left(\frac{x + 1/4}{y + 1/4}\right). \tag{24}$$

For consistency, the expressions for the remaining quantities which are relevant to the analysis that follows must be reviewed in this region of densities:

$$R_1 \simeq K + x + (1 + \frac{2\varepsilon}{\gamma})y \tag{25}$$

where $K \equiv \ln 2 + \dfrac{1}{2} + \dfrac{1}{4}\left(\dfrac{2\varepsilon}{\gamma}\right)$ is a constant. For the eigenvalues,

$$\lambda_1 = \lambda_1(x, y) \simeq -2(x + y)\gamma[1 + \frac{2\varepsilon}{\gamma}(x - y)]$$

$$\lambda_2 = \lambda_2(x, y) \simeq (\tfrac{1}{2} - (x + y))\gamma[1 - \frac{2\varepsilon}{\gamma}(x - y)]. \tag{26}$$

Observe that although the results (25) for $R_1$ and (24) for $R_2$ hold in limit $\dfrac{\varepsilon}{\gamma} \ll 1$ only if both $|x(z,t)|, |y(z,t)| \ll 1/4$, the expressions (26) for $\lambda_1$ and $\lambda_2$ hold valid for any $x$ and $y$ that keep the densities $u(z,t)$ and $c(z,t)$ (22) within the interval $[0, 1]$. We may now use (24), (25) and (26) to examine the behavior of both $x(z,t)$ and $y(z,t)$ through the analysis of the characteristic curves for certain initial and boundary conditions.

## 3.1 Characteristic lines

We intend to examine the behavior of solutions $c(z,t)$ and $u(z,t)$ for all $z \geq 0$ and $t \geq 0$ satisfying the pair of equations (9), for initial conditions (IC) $(z > 0; t = 0)$:

$$c(z, 0) = 1/4$$

$$u(z, 0) = 1/4 + h \tag{27}$$

meaning that $x(z, 0) = h$ and $y(z, 0) = 0$, and boundary conditions (BC) at $z \leq 0; t \geq 0$,

$$c(z \leq 0, t) = 1/4 + y_r \quad \text{i.e.} \quad y(z \leq 0, t) = y_r$$

$$u(z \leq 0, t) = 1/4 + x_r \quad \text{i.e.} \quad x(z \leq 0, t) = x_r. \tag{28}$$

The constants $x_r$, $y_r$, $h$ are such that $|x_r|$, $|y_r|$, $|h| << 1/4$. Here, $h$ represents the excess ($h > 0$) or depletion ($h < 0$) of motor density with respect to the value $1/4$ along the axon at the initial time $t = 0$. The quantities $x_r$ and $y_r$ represent, respectively, the excess or depletion of motors and cargoes densities in the reservoir at $z \leq 0$ and $t \geq 0$.

In the following we present a quantitative analysis of the behavior of the two functions $u(z,t)$ and $c(z,t)$ with respect to time at each point of the domain ($z \geq 0$) using the method of characteristics. We shall restrict this analysis, however, to certain initial and boundary values such that shocks and rarefaction of characteristic lines related to each of the eigenvalues $\lambda_1$ or $\lambda_2$ would not be relevant up to a first approximation in the small parameter $\varepsilon/\gamma$ to figure out solutions within considered time and space domains. As we shall see, this approximation is supported by numerical data. Cases for which it does not hold but may be interesting concerning the transport phenomena of interest here are treated numerically and considered afterward for qualitative analysis as an extension of the perturbative results. A quantitative analysis of the equations (9) for more general initial and boundary values is not on the scope of the present work.

### 3.1.1 Case h > 0

At $t = 0$ and $z \geq 0$,

$$\lambda_1(x(z,0), y(z,0)) = \lambda_1(h,0) \simeq -2h\gamma \left[1 + \frac{2\varepsilon}{\gamma}h\right] < 0 \qquad (a)$$

$$\lambda_2(x(z,0), y(z,0)) = \lambda_2(h,0) \simeq \left(\frac{1}{2} - h\right)\gamma \left[1 - \frac{2\varepsilon}{\gamma}h\right] > 0 \quad (b)$$

(29)

We shall consider $\frac{2\varepsilon}{\gamma}h << 1$. This means that the set of characteristic lines defined as $C(I)$ along which $R_1$ in (25) is conserved have negative slope while the characteristics from the set $C(II)$ along which $R_2$ in (24) is conserved have positive slope. Figure 1 suggests that these features may be extended consistently to the solutions $x(z,t)$ and $y(z,t)$ for all $t \geq 0$. In fact, notice first that the characteristic $z(t) = \lambda_2(h,0)t$ from the set $C(II)$ emerging from $z = 0$ divides the plane into two regions. Any point within the region $z(t) > \lambda_2(h,0)t$ referred to as *Quiet Region (QR)*, is reached by characteristics that belong to both sets $C(I)$ and $C(II)$ emerging at $t = 0$ from all points $z \geq 0$. The points within the region $z(t) < \lambda_2 t$ referred as *Region 1 (Z1),* are also reached from the set of characteristics $C(I)$ emerging at $t = 0$ from the points $z \geq 0$. However, the curves from the set $C(II)$ that reach the points within *Z1* are those emerging from the boundary $z = 0$ at $t > 0$. Thus, while the quantities $x(z, t)$ and $y(z, t)$ are resolved within the *QR* by extending back along both sets of characteristics $C(I)$ and $C(II)$ from any point $(z, t)$ up to the initial conditions, it happens that within *Z1* the corresponding quantities $x(z, t)$ and $y(z, t)$ are resolved extending back from both the initial configuration through $C(I)$ and from the boundary through $C(II)$.

Consider then any point $Q = (z_Q, t_Q)$ at the crossing of two characteristics within the *QR*, one from the set $C(I)$ and the other from the set $C(II)$. Since the initial conditions are such that both densities are constants for $z > 0$, one may write:

$$R_1(z, 0) = K + h = K + x_Q + (1 + \frac{2\varepsilon}{\gamma})y_Q$$

$$R_2(z, 0) = 1 + 4h = \frac{1 + 4x_Q}{1 + 4y_Q}$$

(30)

where we have defined $x_Q \equiv x(z_Q, t_Q)$ and $y_Q \equiv y(z_Q, t_Q)$. From these two equations, one concludes that at any point $Q$ within $QR$, $x_Q = h$ and $y_Q = 0$. This means that the characteristics are straight lines within this region with slopes given in (29).

Consider now the region *Z1* defined by the points $(z, t)$ such that $z \leq \lambda_2(h, 0)t$, $t > 0$. Figure 1 shows two lines that belong to the set $C(I)$ crossing a line from the set $C(II)$ at points $R$ and $S$ within *Z1* and at points $P$ and $Q$ within the *QR*. It follows that

$$R_2(z_R, t_R) = R_2(z_S, t_S)$$

$$R_1(z_P, t_P) = R_1(z_R, t_R) \qquad (31)$$

$$R_1(z_Q, t_Q) = R_1(z_S, t_S)$$

Because $y_Q = y_P$ and $x_Q = x_P$ (*QR*), one concludes that

$$(y_R - y_S)\left(1 + \frac{\varepsilon}{\gamma}\right) = 2(x_S\, y_R - x_R\, y_S) \qquad (32)$$

which has the trivial solutions $y_R = y_S$ if $x_R = x_S$, implying that the characteristics from $C(II)$ are straight lines within the region *Z1*. Now, in order to resolve for the quantities $x_+^* \equiv x(z^*, t^*)$ and $y_+^* \equiv y(z^*, t^*)$ at any point $(z^*, t^*)$ of *Z1*, consider that

$$R_1(0,t) = R_1(z,0)$$

$$R_2(z^*, t^*) = R_2(0,t) \tag{33}$$

$$R_1(z^*, t^*) = R_1(z, 0)$$

Using expressions (24) and (25), one finds that, for the chosen initial and boundary conditions and for $h > 0$, the excess or depletion of cargoes $y_+^* \equiv y(z^*, t^*)$ and motors $x_+^* \equiv x(z^*, t^*)$ at any point $(z^*, t^*)$ within the Z1 are given by:

$$\begin{aligned} y_+^* &= y_r \\ x_+^* &= h - \left(1 + \frac{2\varepsilon}{\gamma}\right) y_r \end{aligned} \tag{34}$$

With regard to the characteristics $C(I)$ within Z1, one must examine the set of equations relating both quantities $R_1(z,t)$ and $R_2(z,t)$ evaluated at any two points $R$ and $T$, as shown:

$$R_1(z_R, t_R) = R_1(z_T, t_T) = R_1(z, 0)$$
$$R_2(z_R, t_R) = R_2(z_T, t_T) = R_2(0, t) \tag{35}$$

The last equality follows from the particular choice of constant boundary conditions as in (28). From these, one may conclude that $x_R = x_T$ and $y_R = y_T$ meaning that both $x$ and $y$ are individually conserved along the characteristics $C(I)$. In turn, this implies that $\lambda_1$ is constant and thus the characteristics of $C(I)$ are also straight lines within Z1. From (34), one concludes that for

$$y_r < \frac{h}{\left(\frac{2\varepsilon}{\gamma}\right)} \tag{36}$$

the condition $x_+^* + y_+^* > 0$ holds assuring that $\lambda_1(x_+^*, y_+^*) < 0$ within Z1, and also that for

$$y_r > \frac{h}{\left(\frac{2\varepsilon}{\gamma}\right)} \left(1 - \frac{1}{2h}\right) \tag{37}$$

condition $1/2 - (x_+^* + y_+^*) > 0$ holds assuring that $\lambda_2(x_+^*, y_+^*) > 0$ within Z1 so that the characteristics from $C(II)$ have positive slope in this region. These conditions give support to the picture outlined in Figure 1.

In the forthcoming examples we show that for the initial conditions (27) with $h > 0$ one can choose values for $x_r$ and $y_r$ for which both $\lambda_1(x, y)$ and $\lambda_2(x, y)$ are approximately conserved within the domain up to the order $\dfrac{2\varepsilon}{\gamma}$. This means that the characteristic lines from each set $C(I)$ and $C(II)$ are nearly parallel to each other. Therefore, in looking for solutions to $u(z, t)$ and $c(z, t)$, shock formation as well as rarefaction regions can be neglected up to this order of approximation. On the basis of the results discussed above one concludes that the system evolves in time exhibiting two regions on the $(z, t)$ plane presenting distinct values for the excess of cargoes and motors densities. Within $QR$, these quantities preserve their initial values whereas within $Z1$, these quantities show also a dependence on the excess of cargoes $y_r$ at the boundary $z = 0$. The speed with which $Z1$ invades $QZ$ is $\lambda_2(x_+^*, y_+^*)$ (29). In the present context this characterizes a wave carrying excess (or depletion) of motors and cargoes that travels towards the positive direction (anterograde transport) along the axon at this speed. We shall see next how this picture may change for $h < 0$.

### 3.1.2 Case h < 0

In this case one notices that at $t = 0$, both $\lambda_1(h, 0)$ and $\lambda_2(h, 0)$ are positive. We shall show that these features are consistent with the solutions $x(z, t)$ and $y(z, t)$ for all times $t > 0$ and space $z \geq 0$ domains, as suggested by Figure 2. In the absence of rarefactions, the characteristics from both sets $C(I)$ and $C(II)$ emerging from $z = 0$ at $t = 0$ divide the plane $z \geq 0$, $t > 0$ into three regions, which will be referred to as the *Quiet Region QR* (upper plane $z(t) > \lambda_2 t$), *intermediate region Z1* ( $\lambda_1 t < z(t) < \lambda_2 t$) and *lower region Z2* ($\lambda_1 t > z(t)$).

Likewise the case $h > 0$, the functions $x(z, t)$ and $y(z, t)$ can be resolved at any point $(z, t)$ within the $QR$ by extending back from $(z, t)$ to the initial values at $t = 0$ along characteristics from both $C(I)$ and $C(II)$ leading to the same equations as in (30). This allows us to conclude that both $C(I)$ and $C(II)$ are straight lines and also that $x_Q \equiv x(z_Q, t_Q) = h$ and $y_Q \equiv y(z_Q, t_Q) = 0$, at any point $Q$ within this region.

The analysis of the solutions within the intermediate region ($Z1$), are performed by choosing any pair of points at the crossing of any two characteristics chosen from the sets $C(I)$ and $C(II)$, say points $R$ and $S$, both at the same characteristic of $C(II)$, as shown in Figure 2. Analogous equations (31) can be set for these two points from which we conclude that the $C(II)$ are straight lines within $Z1$. However, the first of the equations formulated in (33), i.e. $R_1(0, t) = R_1(z, 0)$ does not hold in this case. In order to find the solutions $x_-^*$ and $y_-^*$ to $x$ and $y$ for $h < 0$ at any point $(z^*, t^*)$ of $Z1$, we observe that the set (35) holds also for $h < 0$ yielding $x_R = x_T$ and $y_R = y_T$ at any two points $R$ and $T$ of a $C(I)$. This implies that $\lambda_1$ is also conserved within $Z1$ and thus the characteristics of $C(I)$ are also straight lines within this region. The last two equations in (33) can then be resolved, which yields in the present case,

$$y_-^* = \frac{1}{4}\left[\frac{(1+4h)(1+4y_r)-(1+4x_r)}{\left(1+\frac{2\varepsilon}{\gamma}\right)(1+4y_r)+(1+4x_r)}\right] \tag{38}$$

$$x_-^* = h - \left(1+\frac{2\varepsilon}{\gamma}\right)y_-^*$$

The characteristics of $C(I)$ have positive slope, i.e. $\lambda_1(x_-^*, y_-^*) > 0$ at all points within $Z1$ if $x_-^* + y_-^* = h - \frac{2\varepsilon}{\gamma}y_-^* < 0$. Using the result (38) for $y_-^*$, one shows that this condition is equivalent to:

$$(1+4y_r) > \frac{\left(\frac{2\varepsilon}{\gamma}+4h\right)}{\left(\frac{2\varepsilon}{\gamma}-4h\right)}(1+4x_r). \tag{39}$$

Consider now any point $A = (z_A, t_A)$ within the $Z2$, at the crossing of two characteristics, one from the set $C(I)$ and the other from $C(II)$. The quantities $x(z_A, t_A)$ and $y(z_A, t_A)$ are resolved within the $Z2$ by extending back from $(z_A, t_A)$ along both of these characteristics up to the boundary at $(0,t)$, for any $t \geq 0$. Since the excess densities $x_r$ and $y_r$ at the boundary $z = 0$ are kept at the constant values as time passes these lead to:

$$\begin{aligned}R_1(z_A, t_A) &= R_1(0,t)\\ R_2(z_A, t_A) &= R_2(0,t)\end{aligned} \tag{40}$$

It follows that

$$x_A = x_r \quad \text{and} \quad y_A = y_r \tag{41}$$

at any point $A = (z_A, t_A)$ within $Z2$. Both $C(I)$ and $C(II)$ are straight lines within $Z2$. Moreover, for

$$x_r + y_r < 0 \tag{42}$$

the signs of both slopes do not change with regard to the other regions $QZ$ and $Z1$.

From this analysis one concludes that for $h < 0$, and under absence of shocks or rarefaction of characteristics, the system evolves in time exhibiting three regions on the $(z,t)$ plane each of which presenting, in the most general case, different values for the excess of cargoes and motors.

Within $QZ$, these quantities preserve their initial values. Within $Z1$, these quantities show both a dependence on the initial values and on the values of excess of motors $x_r$ and that of cargoes $y_r$ at the boundary $z = 0$, as shown in (40). Within $Z2$, $x$ and $y$ coincide with their values at the boundaries (41). Under these conditions, the speed at which $Z1$ invades $QZ$ is $\lambda_2(x_-^*, y_-^*)$. The speed at which the region $Z2$ invades $Z1$ is $\lambda_1(x_r, y_r)$.

# 4  Numerical study

The above results for the behavior of both motor and cargo densities achieved analytically for certain initial and boundary conditions are limited to situations for which shocks among characteristic lines and rarefaction regions are either absent or can be neglected within the considered perturbative limit. The numerical simulations performed in parallel to the analysis presented above to study the temporal behavior of $u(z, t)$ and $c(z, t)$ aims to extend the scope of applications attempting to approach more closely the available data on axonal transport. In this Section we briefly discuss some features of the numerical code developed to this end.

The non-linearity of the set of Eqs. (9) can create solutions with both discontinuities and small scale smooth structures. These features together with the expectation that the system develops velocities at very different orders of magnitude, led us to choose, among many numerical methods available, a finite-difference (FD) with a high-resolution shock-capturing (HRSC) scheme, widely used in astrophysical and cosmological codes [21], [22], [23].

The FD method has been chosen since it does not require the solutions of Riemann problems at each interface between computational cells [16] reducing in this way the time costs if compared to the conventional finite volume method (FV). The HRSC methods offer high order of accuracy, sharp descriptions of discontinuities and convergence to the physically correct solution. It has the advantage of treating discontinuous solutions consistently and automatically wherever they appear in the flow [24].

In this work we have implemented our own numerical code using modern FORTRAN with a modular approach. The code uses up to fifth-order reconstruction in the characteristic fields and a local Lax-Friedrichs flux splitting [25]. For a reconstruction scheme, we implement the classic Weighted Essentially Non-Oscillatory (WENO) schemes with third (WENO3) and fifth order (WENO5) [16], [26] and two different improved methods, a third order WENO3p [27] and a fifth order WENO5Z [28]. In addition to the aforementioned references, a brief and practical expla-nation of these methods can be found in Ref. [23]. The time integration of the ODE's obtained from the discretization of Eqs. (9) has been made using a third order Strong Stability-Preserving Runge-Kutta (SSP-RK) scheme [29].

In order to ensure accuracy in the implementation of the code, we first wrote it for the classical Euler equations to run some of the usual validation tests. We then performed a few minor changes to fit the code for our purposes. **Figure 3** depicts the results for the density curves obtained from three different tests and corresponding reconstruction schemes. In the upper left, we show the Sod test [30] evolved up to the time $t_{end} = 0.2$ with 100 cells. This test does not impose significant computational difficulties. Analytical solutions to the equations allows one to check for the accuracy with which the discontinuities are described by the implemented schemes. The blast wave considered in the second test (Figure 3, upper right) has been evolved up to $t_{end} = 0.015$ also with 100 cells. This is a stronger test because the initial profile presents a gradient of five orders of magnitude in pressure. The evident differences in accuracy reached by the different schemes expresses the superiority of fifth order methods. Finally, the results using the iterative blast waves [31] are shown in Figure 3, lower left up to $t_{end} = 0.038$, for 400 cells. These results suggest that the best accuracy among the considered methods is provided by the WENO5Z which solves slightly better for the peaks and valleys than the classic WENO5. Since this problem does not present analytical solutions, we compare the results for this 400 cells with the outcomes of a more accurate calculation, referred to as "exact" and obtained using WENO5 with 2000 cells.

All the above tests suggest that the methods have been well implemented for the classical one-dimensional Euler set composed of three equations for four unknowns (mass density, velocity, pressure´and energy density) which must be complemented with an equation of state (EoS) providing in this way a relationship among the unknowns. The equations considered in the present work compose a closed set with two equations and two unknowns (the motor and cargo densities), making it possible to perform some simplifications in respect to the application to the Euler problem, in order to improve the computational costs. An additional test applied to the Linearized Gas Dynamics was then performed [32] . This corresponds to a classical Riemann Problem with a fluid at rest and density equal to 1.0 on the left side and 0.5 at its right. We then investigate the evolution of the density and velocity profiles of an ideal gas with transmissive boundary conditions. In **Figure 3, lower right** it is shown the results for density and velocity profiles obtained both analytically and also by means of a simulation run for 100 cells using a WENO5Z scheme. The two solutions are in good agreement with each other, even near discontinuities.

# 5   Discussion of the results: theory and simulations

The analytical expressions obtained in the preceeding section for motor and cargo densities are restricted to the perturbative regime for which the parameter $\varepsilon/\gamma \ll 1$. In addition, the initial and boundary values have been chosen as small differences $x$ and $y$ with respect to the densities at the defined background corresponding to the value $1/4$ in Eq. (22). Nonetheless, the results obtained for the two eigenvalues $\lambda_1$ and $\lambda_2$ in Eq. (15), are only restricted by the smallness of $\varepsilon/\gamma$. The simplified expressions (26) or (29) for the particular initial conditions expressed in terms of the parameter $h$, resulted from direct substitution of the defined quantities $x$ and $y$. This suggests that the speed of the interface separating the $QZ$ from $Z_1$ that in the absence of shocks is given by $v_c = \lambda_2(x^*, y^*) \simeq \lambda_2(h, 0)$, can be analyzed for any $h$ within the interval $-1/4 < h < 3/4$. This allows us to predict that, for $|h| \ll 1$ the speed of the interface is close to its maximum value $\lambda_2^{\max} = 0.5$ whereas for $h \sim 0.5$ the speed $\lambda_2(h, 0)$ of the referred interface shall attain very small values. Therefore, although the expressions derived above for $\lambda_1$ and $\lambda_2$ in terms of the excesses $x$ and $y$ do not provide exact results for the speed of the interfaces, especially in cases for which shock and rarefaction of characteristics are determinants to the dynamics, these expressions function as important guides to direct our numerical experiments in order to scrutinize the parameter space and investigate the diverse behaviors of the quantities of interest. The time evolution of motor and cargo densities for small $|h|$ are expected to display very different profiles compared to those produced for $h \sim 0.5$. Although one cannot predict based on of the above calculations what would be the fates of the motor and cargo density profiles at these relatively high values of $h$, one can indeed envisage that the interface separating the $QZ$ from its neighboring region, might travel at extremely low speeds, if compared to the corresponding speeds at small values of $|h|$.

The following examples support these expectations. Cases (1)-(6) shown in Table 1 illustrate the behavior of the excess of cargoes and motors for $|h| \ll 1$ at the indicated initial $x_0 = h; y_0 = 0$ and boundary values $x_r = x(0, t)$ and $y_r = y(0, t)$. It is also depicted the speed of the corresponding interface between $QZ$ and $Z_1$. For these, the motor hopping rates are fixed at $p = 0.9$ (unloaded) and $q = 0.85$ (loaded). Although other possibilities could be validated, these choices keep the expansion parameter at small values $\dfrac{2\varepsilon}{\gamma} = 0.057$ . The cargo hopping rates between neighbor motors are fixed at $f = g = 0.2$. Fixing $f = g$ leads to $\eta = \theta = 1$ (10) that is consistent with the development in Section **2**. The choices for $h$ and boundary values $x_r$ and $y_r$ in each example and the corresponding numerical $v_{num}$ and analytical $v_c$ values obtained using the results of preceding Section are compiled in Table1. $v_c = \lambda_2(x^*, y^*)$ characterizes the velocity of sharp interfaces between the regions $QR$ and $Z1$.

**Table1 -** Analytical and numerical results for density wave profiles and accompanying velocities within the perturbative regime.

The results for $x^*$ and $y^*$ evaluated with the help of expressions (34) for $h > 0$ or (38) for $h < 0$, and the corresponding $x^*_{num}$ and $y^*_{num}$ obtained by numerical simulation for the excess of motors and cargoes within region $Z_1$ are also shown in Table1. In **Figure 4** there are represented the wave density profiles for cases (2) (Fig. 4a) and (3) (Fig. 4b) at different instant of times, as indicated. These are chosen from examples at typical boundary values $x_r$ and $y_r$ for which the system presents at initial times either slight depletion of motors $h < 0$ or a slight excess $h > 0$ with respect to the background at $u = 1/4$. The corresponding differences in the average cargo velocities reflect the effects due to motor jamming. The analytical results achieved for the velocities and density profiles are in fact very well represented by their corresponding numerical results. This supports the idea of using these expressions to guide the choice of parameters in order to investigate the behavior of the quantities of interest for the axon transport problem within regions away from the perturbative limits.

Results in Table 2 for speed of the wave front $v_q^p$ at which the $QZ$ is invaded as time passes have been obtained by numerical simulation, as described above. These are indexed by the values for the parameters $p$ and $q$ as shown in each column. We use $(p = 0.9,\ q = 0.85)$; $(p = 0.7,\ q = 0.5)$; $(p = 0.2,\ q = 0.1)$ and $(p = 0.3,\ q = 0.25)$. These results for $v_q^p$ cannot be reproduced by the analytical expressions derived above because the initial and boundary values have been fixed, respectively, at $h = 0.5$, $x_r = -0.1$ and $y_r = 0.1$ which do not fully satisfy the perturbative conditions. Nonetheless, we expect in these cases that the pulse fixed by the boundary values advancing as a wave profile over $Z1$ in each case do that at very small speeds, as suggested by the expression for $\lambda_2$ in (26). The results below illustrating the dependence of $v_q^p$ on the microscopic parameters $f$ and $g$ corroborate these predictions: for large $h$, i.e. $h = 0.5$ the numerical values depicted for $v_q^p$ are typically orders of magnitude less than the values $v_q^p$ obtained for $|h| = 0.05$. **Figure 5** illustrates the corresponding cargo and motor wave density profiles at different instants of time obtained numerically for two situations, case (5) Fig.5(a), and case (6) Fig.5(b), taken from Table 2. For chosen values of parameters $f$ and $g$, we compare the profiles of $h = 0.5$ and $h = -0.05$ at fixed $p = 0.95$ and $q = 0.85$. These results suggest that to the relatively large differences depicted by cargo velocities there are associated large differences on the wave profile, as it is observed in experimental data [2].

**Table 2** - Cargo average velocities $v_q^p$ for motor hopping parameters $p$ (unloaded), $q$ (loaded) and excess on the track $h$ at initial times as indicated, within the non-perturbative regime.

Table 2 depicts typical behaviors of the average speed $v_q^p$ at which cargoes from the boundary at $z = 0$ advance over the $QZ$. We observe that:

(i) $v_q^p$ decreases as the excess of motors $h$ on the axon increases. As $h$ reaches 0.5, keeping the remaining parameters at fixed values $v_q^p$ may decrease to such small values up to 3 orders of magnitude smaller than those reached at $|h| = 0.05$. This is due to an increase of motor traffic jam that hampers loaded-motor motion through process $q$. In this case, the important mechanism for driving cargo movement is cargo hopping regulated by rates $g$ (forwards) and $f$ (backwards). In turn, this may explain the existence of variations in the velocity of slow components that travel on axons known as slow component-a (Sca) and slow component-b (Scb), whose speed may differ from each other up to one order of magnitude. These variations are compatible with results in Table 2 as parameters $g$ and $f$ change at high values of $h$.

(ii) If $f = g$ and for $h$ not too large, $p$ and $q$ become the dominant processes to drive cargoes. The relatively high values of $v_q^p$ reached at $|h| = 0.05$ (Table 2) reflect the success of mechanisms $p$ and $q$ to drive the cargoes in the absence of heavy traffic jam (congestion).

(iii) In general, $v_q^p$ increases with parameters $p$ and $q$, as expected. However, the effects of increasing $p$ (unloaded motors) and $q$ (loaded motors) in a situation of high motor density should not result in significant increasing of velocities since traffic jam is the determinant effect in this case, as mentioned. Consistently with this, we expect that the effect of increasing $v_q^p$ with $p$ and/or $q$ shall be noticeable at low motor densities, as observed in the results depicted above.

(iv) The effects on $v_q^p$ of changing cargo hopping rates from $g < f$ to $g > f$ are such that $v_q^p$ increases at high motor excess $h = 0.5$ but decreases at small $|h| = 0.05$. For discussing this effect one should notice that the dominant process for driving cargoes at high densities is hopping controlled by the rates $f$ and $g$. We should then expect that $v_q^p$ increases by increasing the forward hopping rate $g$ with respect to $f$. On the other hand, one observes in Table 2 that at low motor densities $|h| = 0.05$ the forward speed $v_q^p$ does not increase with $g$. On the contrary, it increases as the backward cargo hopping rate $f$ increases with respect to $g$. This can be understood since at low densities, the clusters of motors eventually assembled are likely to appear isolated from the others so that for $g > f$ the loads would end up stacked in front of each cluster. In this case, and for relative small $q$, each cluster would take a long time to reach the back end of the next cluster. On the other hand, for $g < f$ the cargoes would be more prone hopping backward to accumulate at the back end of each cluster until another cluster, led by its front motors through process $p$, would join the former providing in this way extra home for process $f$ to continue. Accordingly, the accumulation of cargoes at the back end of the new cluster of motors assembled in this way by the union of the previous two would be the source of shock waves driving cargoes to the forward direction.

In this context, we want to argue about a possible connection between the results achieved here and those reported from experiments performed to investigate the effects of Kinesin-Binding Proteins (KBP) on the transport of particles along microtubules [13]. In that study, one investigates the effects on particle mobility under variations of the amount of KBP in the considered systems. Effects of increasing KBP have been correlated to a decrease of KIF1A motors attached to the microtubules, and to an increase in the average velocities of the remaining moving KIF1A motors after the introduction of KBP. On the contrary, it is observed in another experimental series that average velocities of Kinesin based mobile Rab vesicles (cargoes) decrease at increasing KBP (i.e., decreasing the amount of motor attached).

Figure 6 depicts the predictions of the model for the behavior of average velocities, under variations of the excess $h$ for certain choices of parameters, as indicated. Both velocities present a unique global maximum at $h = h_{\text{max}}$, that depends on the choices for $f$ and $g$. This behavior can be understood as resulting from two effects: increasing the excess of motors at low densities ($h < h_{\text{max}}$) would contribute to drive the cargoes that are released from the reservoir. On the other hand, effects due to traffic jam would prevail at large motor densities ($h > h_{\text{max}}$), as discussed above. We may then suggest that the data in [13] might be understood as resulting from a balance between these two effects under the condition that the related experiments explore different regions of motor densities.

# 6 Conclusions and additional remarks

Our model describes axon transport of different particles in a unified mode as these are conveyed directly by molecular motors, in the absence of intermediates i.e. vesicles or any other *a priori* moving structure [4], [5]. The idea is that the cargoes are able to detach from molecular motors available on the axons and attach directly to neighbor motors at certain probability rates. This hopping mechanism is favored under situations of a heavy traffic jam at which the motors assemble into large clusters. We propose that such clusters may supply the need for the alleged intermediate carriers. The model accounts also for processes leading to direct movement of cargoes attached to motors if allowed by local jamming conditions. These features are able to predict quantitatively the observed differences between the velocities of the particles conveyed in axons at slow and fast modes, essentially as a consequence of changing the density of motors available because it is the density of motors that regulate the traffic jam. Moreover, the model suggests that the stochastic mechanism of cargo hopping from motor to motor can also explain the differences between the two components at slow rates (Sca and Scb) by tuning the corresponding hopping parameters $f$ and $g$.

In order to be able to extract these properties from the model and describe phenomenological aspects of these systems concerning cargo velocities, we have based our analysis on particular solutions to the pair of coupled equations of the Burgers type describing the dynamics of defined densities of motors and cargoes. These equations resulted from a mean-field approach of the original stochastic description. We use a perturbative scheme for certain initial and boundary values considering the particular choice for which $f = g$. To our knowledge, the analytical solutions achieved in this way to the system of quasi-linear PDE´s are new in the literature. In turn, these solutions pointed out to the regions of parameters, extending far beyond the perturbative limits to be investigated numerically in order to obtain information about the system that may relate to the existing data. Such interplay between numerical and analytical approaches was shown to be crucial in exploring the properties of interest.

It is worth mentioning that the description of interacting particle system considered here allows for studying the effects produced by one kind of carrier motor and one kind of cargo at a time. Different motors and cargoes should be taken into account by a separate dynamics characterized by some specific set of parameters. We have presented results for motors that have been chosen possessing anterograde (plus-ended) bias. Our predictions suggest that depending on the initial conditions, cargoes normally recognized as slow components (either Sc**a** or Sc**b**) may also be con-veyed in fast-moving mode as already observed, for example, in studies exploring the dynamics of synapsin proteins [33]. Accordingly, what would determine if a set of cargoes move in a fast or slow mode is a combination of factors comprising their ability to interact with motors and the quantity of motor available on the axon. In a forthcoming study, we shall consider this same kind of description attempting to unravel the corresponding pulse-like wave profiles observed in axon transport since these present distinct and typical features for slow and fast components.

# 7 Acknowledgements

We wish to thank the faculty members of the Laboratório de Automação e Controle (LAC) from Escola Politécnica da Universidade de São Paulo (Poli-USP) for their criticisms and helpful suggestions. VRM acknowledges the financial support provided by Universidade de São Paulo (USP).

# 8 Figure Caption

- Figure 1 - Characteristic lines for the system (9) with initial and boundary conditions as specified in (27) and (28), $h > 0$.

- Figure 2 - Characteristic lines for the system (9) with initial and boundary conditions as specified in (27) and (28), $h < 0$.

- Figure 3 - Density profiles for numerical tests validation. *Upper left*: Sod's problem. *Upper Right*: Blast Wave Problem. *Lower left*: Interactive Blast Waves. *Lower right*: Linearized Gas.

- Figure 4 - Time evolution of motor (gray) and cargo (black) density profiles for illustrative examples taken from Table 1. (a) - case (2) ; (b) - case (3), for the initial cargo occupation at $c_0 = 0.25$ and reservoir excesses (28) at $x_r = -0.10$ and $y_r = 0.10$. For the scale factor we use $\xi = 1$ what sets $\alpha = p$ and $\beta = q$ (10).

- Figure 5 - Time evolution of motor (gray) and cargo (black) density profiles for illustrative examples taken from Table 2, for $p = 0.90$ and $q = 0.85$. (a) $h = 0.5$ case (5) ; (b) $h = -0.05$ case (6), for the initial and reservoir occupations, and scale factor set as in Figure 4.

- Figure 6 - Variation of the average (a) cargo and (b) motor velocities with the excess motor density $h$, at $c_0 = 0.20$; $p = 0.90$ and $q = 0.85$, for $f = g = 0.2$ (cross); $f = 0.2, g = 0.4$ (plus); $f = 0.4, g = 0.2$ (dot).

**Figure 1**

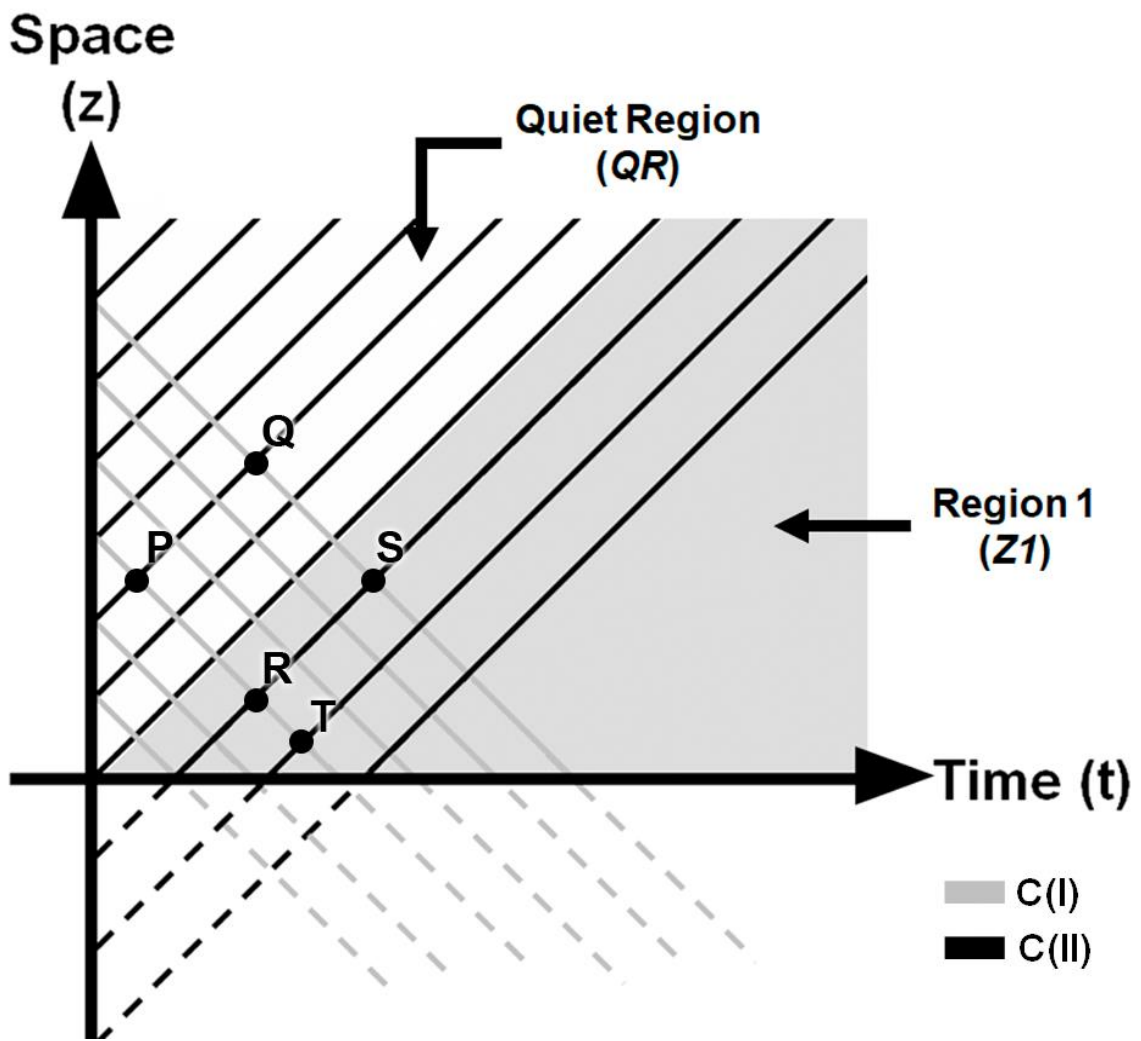

**Figure 2**

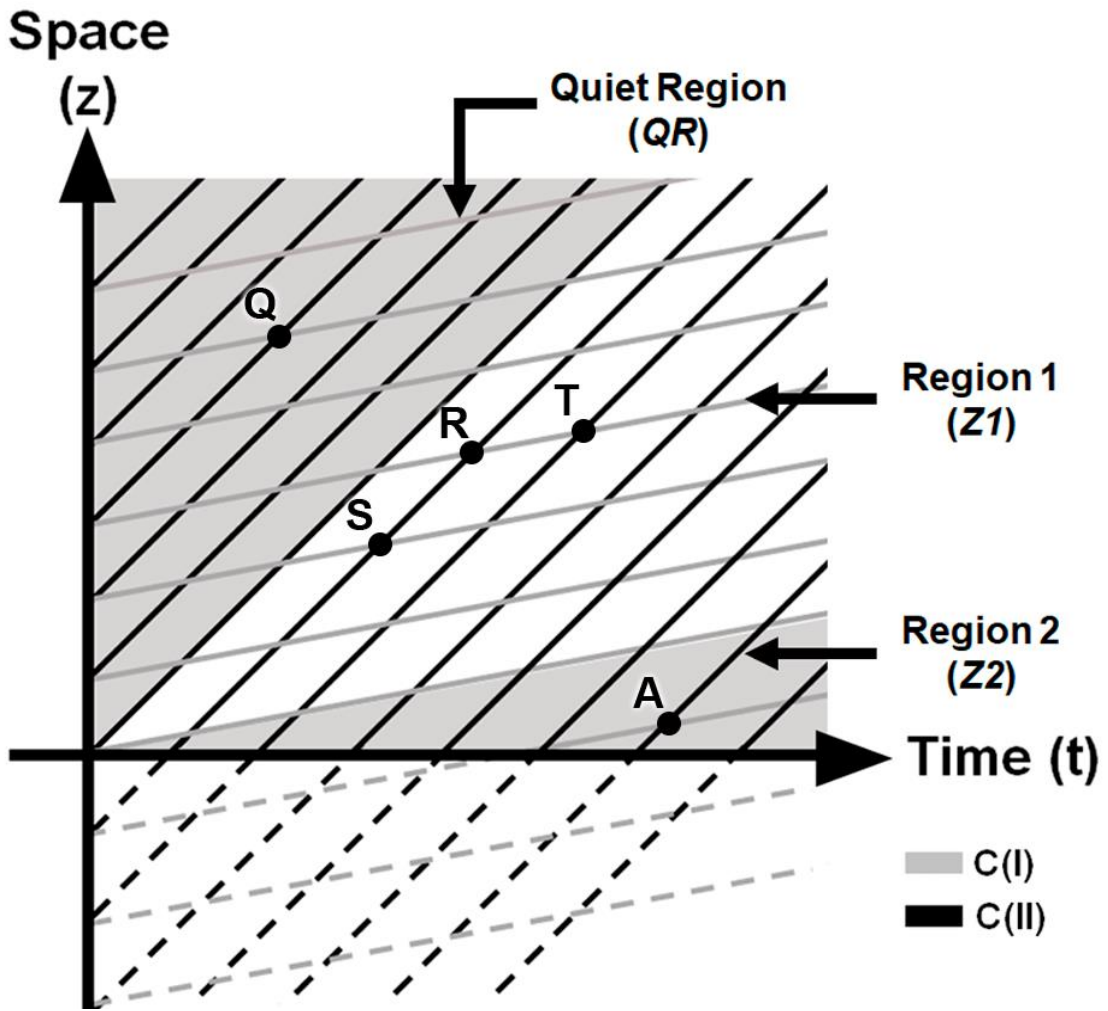

**Figure 3**

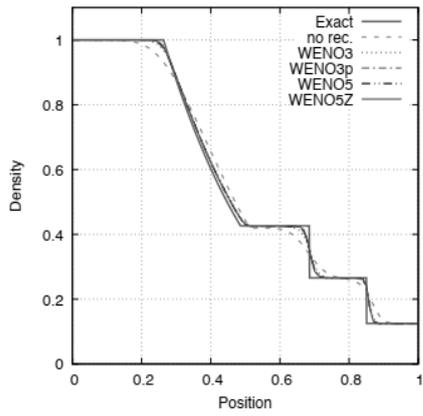
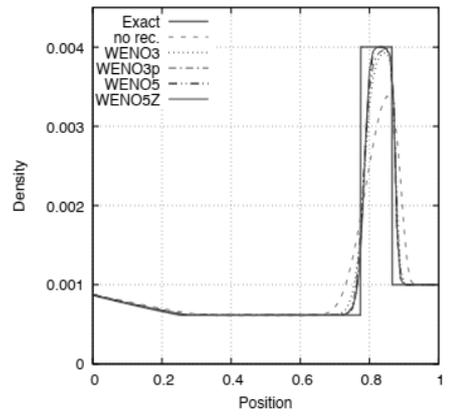
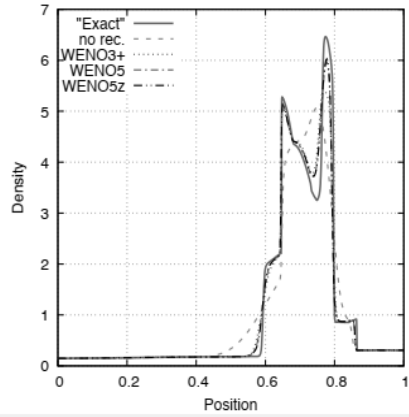
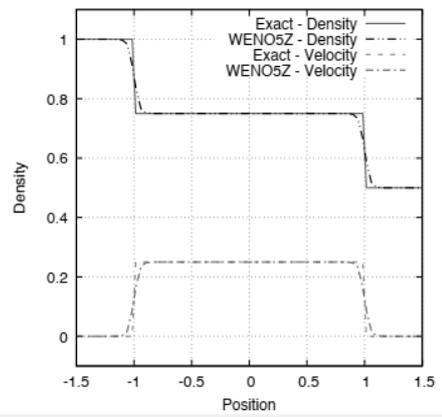

**Figure 4(a)**

t = 0 s

t = 47,54 s

t = 63,74 s

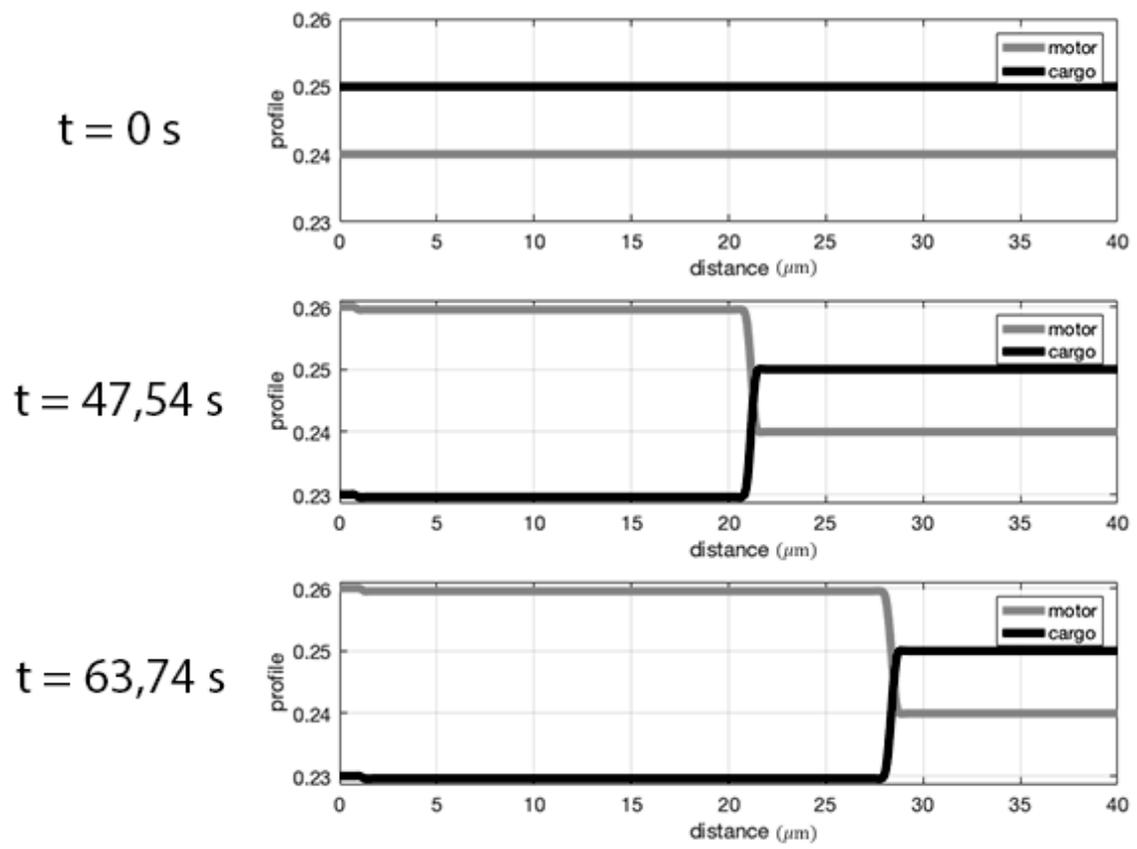

**Figure 4(b)**

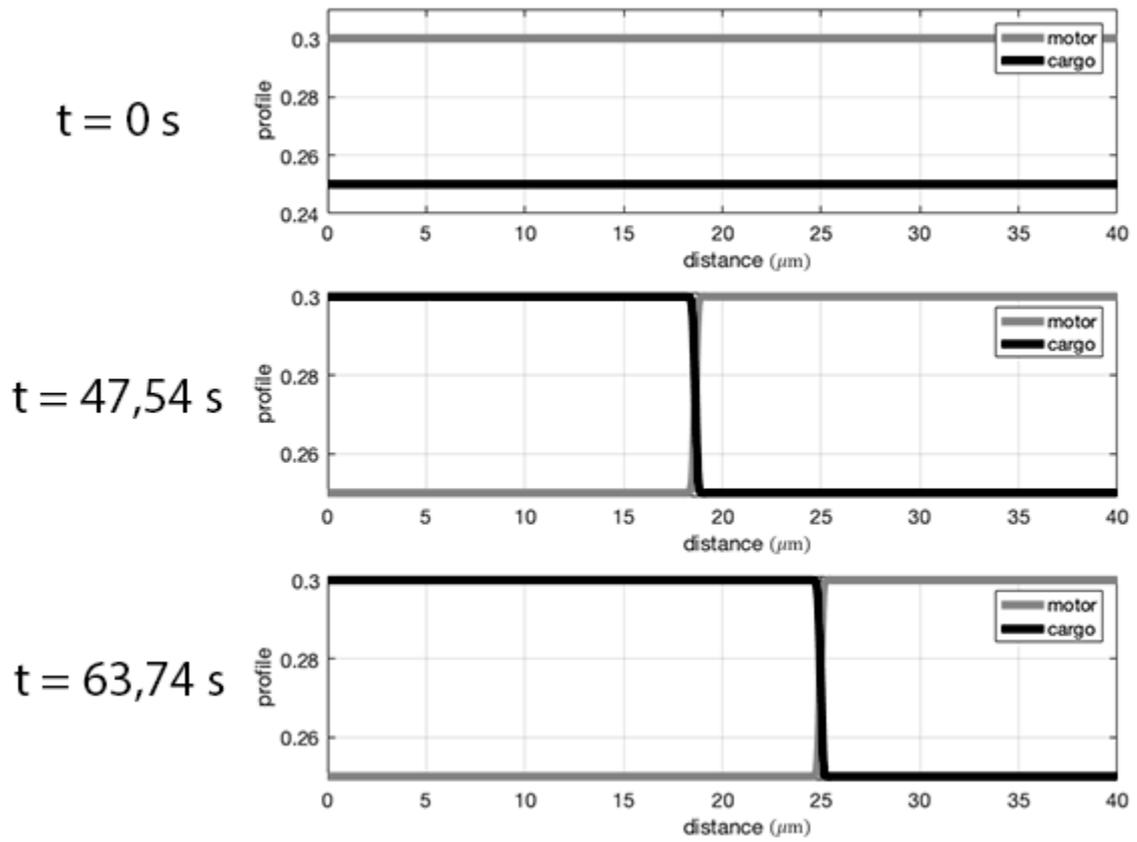

**Figure 5(a)**

t = 0 s

t = 47,54 s

t = 63,74 s

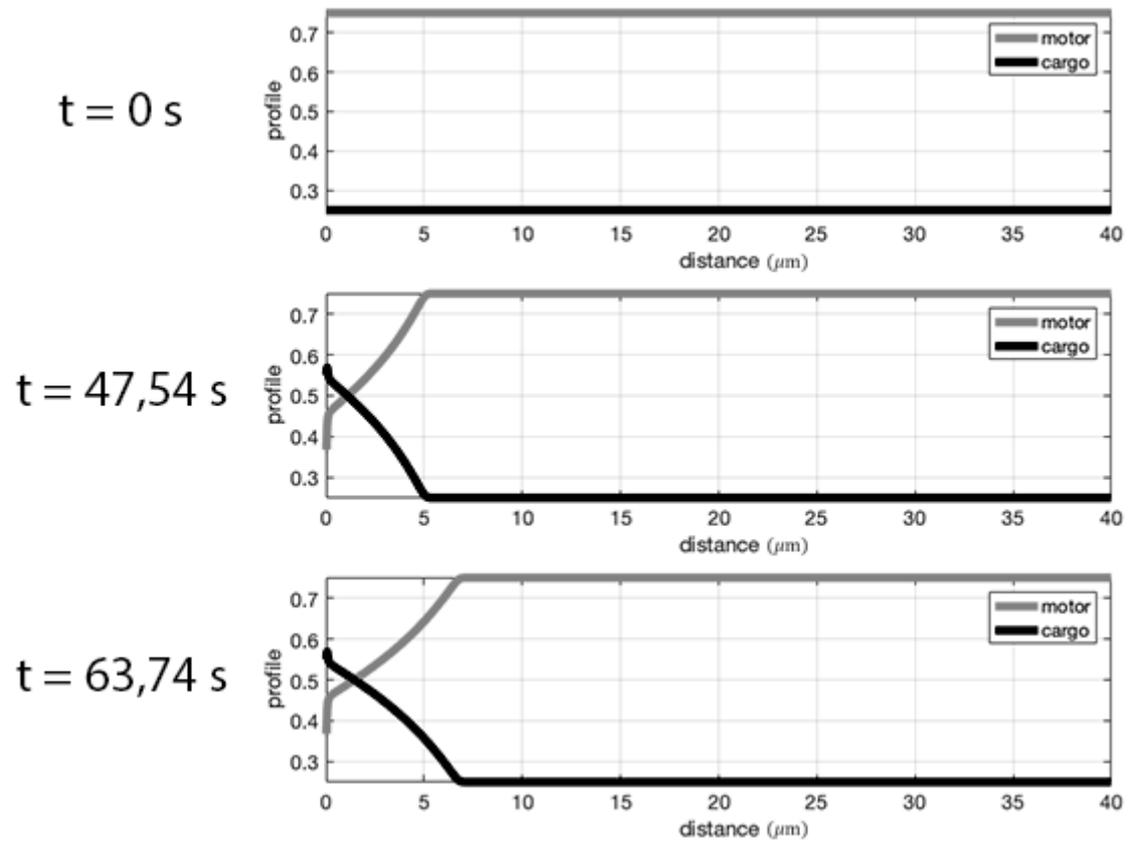

**Figure 5(b)**

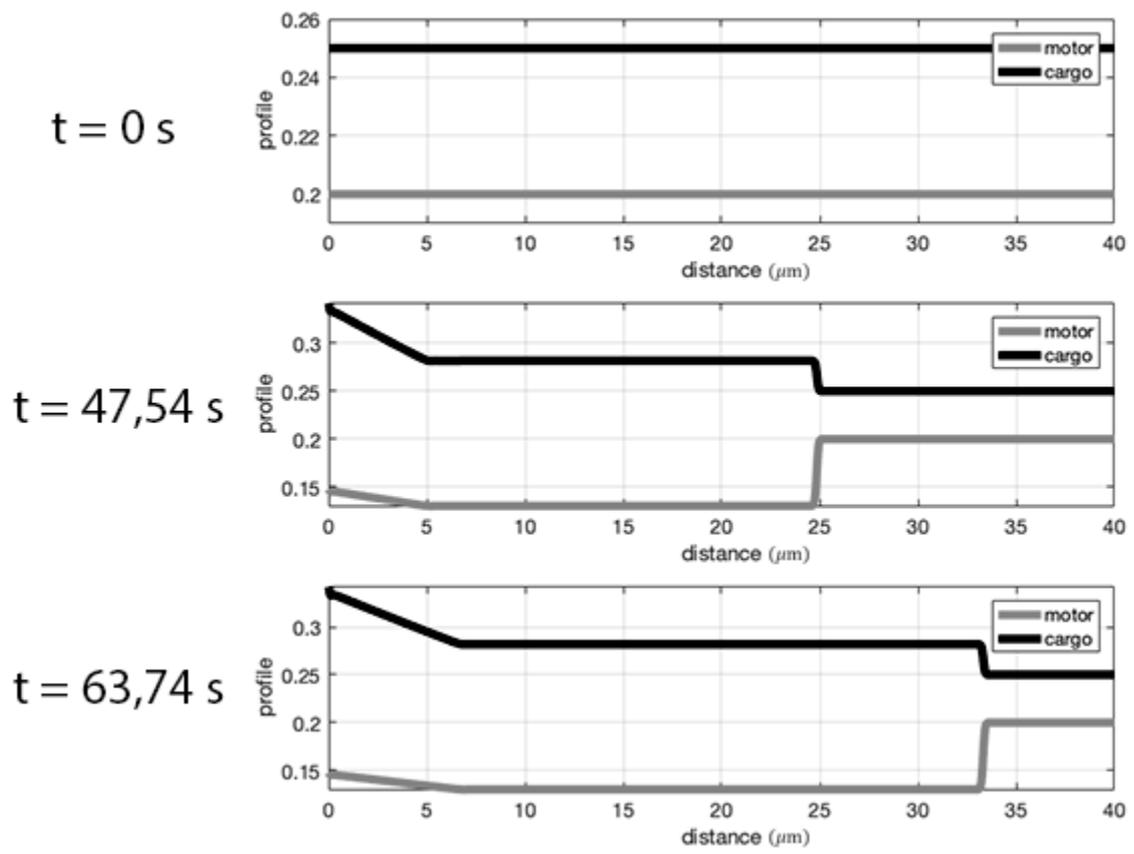

**Figure 6(a)**

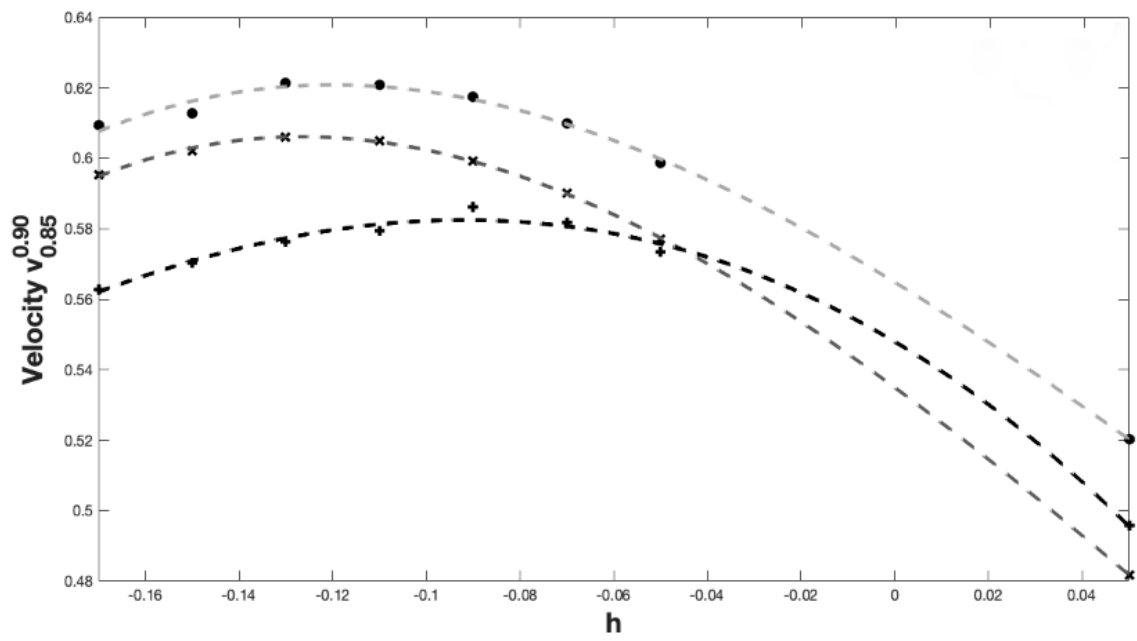

**Figure 6(b)**

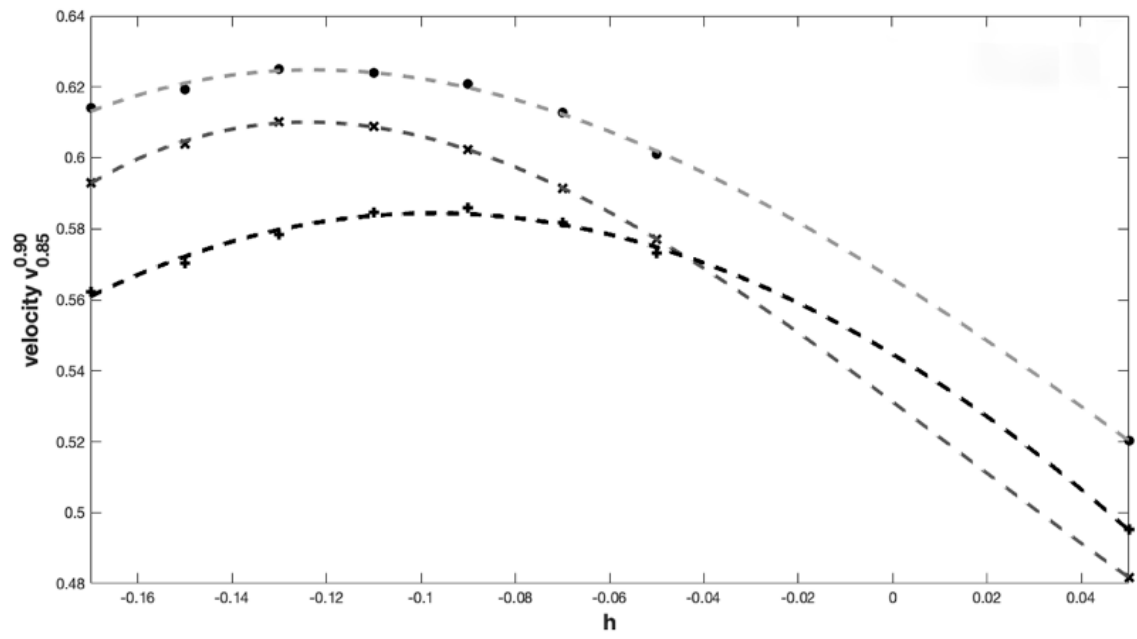

**Table 1**

| Case | $h$ | $x_r$ | $y_r$ | $v_c$ ($\mu m/s$) | $v_{num}$ ($\mu m/s$) | $x^*$ | $y^*$ | $x^*_{num}$ | $y^*_{num}$ |
|------|-------|--------|--------|-------|-------|--------|--------|--------|--------|
| (1)  | 0.01  | 0.02   | −0.01  | 0.429 | 0.434 | 0.020  | −0.010 | 0.020  | −0.010 |
| (2)  | −0.01 | 0.01   | −0.02  | 0.446 | 0.451 | 0.009  | −0.019 | 0.009  | −0.021 |
| (3)  | 0.05  | 0      | 0.05   | 0.394 | 0.396 | −0.03  | 0.050  | 0      | 0.050  |
| (4)  | −0.05 | −0.025 | −0.025 | 0.482 | 0.486 | −0.024 | −0.024 | −0.023 | −0.023 |
| (5)  | 0.01  | 0      | 0.01   | 0.429 | 0.431 | 0      | 0.010  | 0      | 0.010  |
| (6)  | −0.01 | 0.02   | −0.03  | 0.446 | 0.451 | 0.021  | −0.021 | 0.019  | −0.031 |

**Table 2**

| Case | $f$ | $g$ | $h = 0.50$ | | | | $h = 0.05$ | | | | $h = -0.05$ | | | |
|---|---|---|---|---|---|---|---|---|---|---|---|---|---|---|
| | | | $v_{0.85}^{0.90}$ ($\mu m/s$) | $v_{0.25}^{0.90}$ ($\mu m/s$) | $v_{0.50}^{0.70}$ ($\mu m/s$) | $v_{0.25}^{0.30}$ ($\mu m/s$) | $v_{0.85}^{0.90}$ ($\mu m/s$) | $v_{0.25}^{0.90}$ ($\mu m/s$) | $v_{0.50}^{0.70}$ ($\mu m/s$) | $v_{0.25}^{0.30}$ ($\mu m/s$) | $v_{0.85}^{0.90}$ ($\mu m/s$) | $v_{0.25}^{0.90}$ ($\mu m/s$) | $v_{0.50}^{0.70}$ ($\mu m/s$) | $v_{0.25}^{0.30}$ ($\mu m/s$) |
| (1) | 0.01 | 0.01 | 0.00056 | 0.00056 | 0.00056 | 0.00056 | 0.398 | 0.120 | 0.0276 | 0.127 | 0.489 | 0.334 | 0.356 | 0.160 |
| (2) | 0.01 | 0.20 | 0.103 | 0.074 | 0.092 | 0.079 | 0.415 | 0.105 | 0.259 | 0.112 | 0.488 | 0.191 | 0.319 | 0.133 |
| (3) | 0.20 | 0.01 | 0.0082 | 0.0056 | 0.0056 | 0.0243 | 0.436 | 0.331 | 0.331 | 0.152 | 0.518 | 0.444 | 0.400 | 0.133 |
| (4) | 0.20 | 0.20 | 0.00056 | 0.00056 | 0.00056 | 0.00056 | 0.398 | 0.120 | 0.276 | 0.127 | 0.489 | 0.344 | 0.356 | 0.160 |
| (5) | 0.20 | 0.40 | 0.090 | 0.077 | 0.094 | 0.081 | 0.416 | 0.098 | 0.259 | 0.108 | 0.488 | 0.183 | 0.319 | 0.125 |
| (6) | 0.40 | 0.20 | 0.009 | 0.00056 | 0.007 | 0.025 | 0.438 | 0.336 | 0.334 | 0.150 | 0.519 | 0.448 | 0.400 | 0.174 |